\documentclass[aps,pre,twocolumn,superscriptaddress,showpacs]{revtex4}
\usepackage{amssymb}
\usepackage{epsfig}
\usepackage{amsmath}
\usepackage{times}

\begin{document}

\title{Slow dynamics of Zero Range Process in the Framework of Traps Model}

\author{Kai Qi}
\affiliation{Web Sciences Center, University of Electronic Science and Technology of China,
Chengdu 610054, People's Republic China}
\author{Ming Tang}
\email{tangminghuang521@hotmail.com}\affiliation{Web Sciences Center, University of Electronic Science and Technology of China,
Chengdu 610054, People's Republic China}
\author{Aixiang Cui}
\affiliation{Web Sciences Center, University of Electronic Science and Technology of China,
Chengdu 610054, People's Republic China}
\author{Yan Fu}
\affiliation{Web Sciences Center, University of Electronic Science and Technology of China,
Chengdu 610054, People's Republic China}

\date{\today}

\begin{abstract}

The relaxation dynamics of zero range process~(ZRP)~has always been an interesting problem.
In this study, we set up the relationship between ZRP and traps model, and investigate
the slow dynamics of ZRP in the framework of traps model.
Through statistical quantities such as the average rest time, the particle
distribution, the two-time correlation function and the average escape time,
we find that the particle interaction, especially the resulted condensation,
can significantly influence the dynamics. In the stationary state,
both the average rest time and the average escape time
caused by the attraction among particles are obtained analytically.
In the transient state, a hierarchical nature of the aging dynamics
is revealed by both simulations and scaling analysis. Moreover, by comparing the
particle diffusion in both the transient state and the stationary state,
we find that the closer ZRP systems approach the stationary state,
the more slowly particles diffuse.

\end{abstract}

 \maketitle

The investigation of diffusion process plays a
very important role in exploring the structures of systems and
uncovering the physical mechanisms of dynamics in particle
systems~\cite{Hughes,fpp,AH00,Woess}. Recently, with the booming research results in the field
of complex networks, particle diffusion in complex networks has
once again become a hot problem~\cite{Adamic,yang,Sood,Jasch:2001,
Almaas:2003, Stauffer:2005, nohrieger, Gallos:2004, Kittas:2008,
Bollt, Baronchelli:2006, Juhasz:2008,
Baronchelli:2008,Zlatanov:2009,Tejedor:2009,Baronchelli:2010}. For example, Noh \emph{et al.} studied random walks in
scale-free (SF) networks and found that the ratio between a
coordination number and a characteristic relaxation time for each
node essentially determines the MFPT~\cite{nohrieger}.

A common feature of above studies is that there is no
interaction among particles at each node.
Actually, particle interaction is unavoidable and can be found in
many real networks where it plays an important role in the dynamical
processes in networks. One simple way to introduce
the interactions among stochastic particles is so-called zero range
process (ZRP), which has been recently adopted to investigate
particle condensation in complex networks~\cite{Noh:2005a,Noh:2005b,Tang:2006,Waclaw:2007,Tang:2009}. It is shown that this condensation
transition appears in a number of unexpected contexts such as wealth
condensation in macroeconomies~\cite{Burda:2002}, jamming in
traffic~\cite{Evans:1996,Chowdhury:2000,Zhu:2007,Shen:2010}, coalescence in granular
systems~\cite{Eggers:1999,Meer:2004}, and gelation in
networks~\cite{Krapivsky:2000,Bianconi:2001}. To understand the influence of ZRP interaction on the particle diffusion,
we made an important first step to study another aspect of ZRP in scale-free
networks, i.e., the diffusion features~\cite{Tang:2010}. We found
that the statistical quantities of diffusion can be significantly reduced by the condensation
and can be figured out by the rest time of a particle staying at a node. In addition
to these statistical quantities in the stationary state, however, the relaxation dynamics of ZRP are fully overlooked.

Recently, a traps model with interaction, in which particles are attracted
at each node~(i.e.~local minima)~with the potential-energy
landscape~\cite{Bouchaud:1992}, was put forward to study the connection
between the network of the potential-energy
landscape and the glassy dynamics~\cite{Baronchelli:2009}. At low temperatures, the relation between the energy and the
degree of a minimum can result in the slow dynamics of glassy systems.
This provided a systematic integration of tools and concepts to investigate
how network structures impact the particle diffusion. In this letter,
we will try to set up the relationship between ZRP and traps model, and then investigate
the slow dynamics in the ZRP interaction in the framework of traps model.

Firstly, we briefly introduce the ZRP model. In complex networks,
ZRP interaction means that a particle interacts only with other particles
staying at the same node. Suppose $N$ particles are randomly put in a network with $L$ nodes and
each node $i$ can be occupied by any integer number of particles
$n_i=0,1,2,\cdots,N$. Due to interaction, usually only part of the
particles at a node can jump out and hop into its neighboring nodes,
which makes the particles be redistributed among all nodes.
We divide the diffusion process of particles
into two steps. In the first step, some
particles at node $i$ will jump out because of the interaction
among the particles. Suppose a particle at node~$i$~jumps out at the rate
$p(n_i)=n_i^{\delta}$ with $\delta\in
[0,1]$~\cite{Noh:2005a,Noh:2005b,Tang:2006}, where the parameter
$\delta$ can be used to characterize the nature of interaction among
particles at the node. In the second step,
each jumping particle hops from the node $i$
to one of its neighbors $j$ randomly.

In the stationary state, there is a condensation threshold
$\delta_c=1/(\gamma-1)$, and a finite fraction of the total particles will be
condensed to the hubs for~$\delta<\delta_c$~\cite{Noh:2005a,Noh:2005b,Tang:2006}.
Generally, the particle density $\rho=N/L$. In the mean field approach, the description of $n_i$ for each
node is replaced with the mean occupation number $m_k$ for the
nodes with the same degree $k$, i.e., $m_k$ is the average of all
the $n_i$ at the nodes with degree $k$. Hence $m_k$ is not
necessarily an integer. In this
framework, it is shown that the number of mean occupation particle
in the stationary state is~\cite{Tang:2006}
\begin{eqnarray}\label{eq:occupation}
m_k&=&k/k_c, k< k_c;\nonumber \\
m_k&=&(k/k_c)^{1/\delta}, k\geq k_c,
\end{eqnarray}
where the crossover degree $k_c$ denotes the degree for $m_k=1$, and is given by~\cite{Tang:2010}
\begin{equation}\label{eq:kc-delta1}
k_c \simeq \left\{\begin{array}{l} {[\frac{A}{\rho(-\gamma+1+1/\delta)}]}^{\delta}
k_{max}^{1-\delta/\delta_c}, \quad for \quad \delta < \delta_c;\\ {[\frac{A}{\rho}]}^{\delta}[\ln
k_{max}]^{\delta_c}
, \quad for \quad \delta=\delta_c.
\end{array} \right.
\end{equation}
For $\delta > \delta_c$, there is no condensation, and the crossover degree $k_c$ is given by
\begin{equation}\label{eq:kc-delta2}
\rho \simeq
\frac{Ak_c^{-1}}{\gamma-2}(k_0^{-\gamma+2}-k_c^{-\gamma+2})+
\frac{Ak_c^{-\gamma+1}}{\gamma-1-1/\delta}.
\end{equation}

To gain further insight into the relaxation dynamics of the ZRP systems~\cite{Tang:2010}, we first
set up the relationship between ZRP and traps model.
Treating each node as a trap with temperature $T$ and depth $E_k$, we
obtain a trapping network where both $T$ and $E_k$ can be determined by
the jumping rate $p(m)$. As pointed out in our previous work~\cite{Tang:2010}, the jumping rate has different
expressions
\begin{equation}\label{eq:jumping rate}
p(m_k) = \left\{\begin{array}{l} {m_k=k/k_c}
, \quad for \quad k < k_c; \\
 {m_k^{\delta}=k/k_c}, \quad for \quad k \geq k_c.
\end{array} \right.
\end{equation}
That is, $p(m_k)$ is proportional to the degree $k$ and has the
same expression for all degree $k$. As the jumping rate $p(m)$
represents the number of particles jumping out of a given node per
unit time, the rest time $\tau_k$ is $m_k/p(m_k)$, which is given by
\begin{equation}\label{eq:trapping rate}
\tau_k = \left\{\begin{array}{l} {1}
, \quad for \quad k < k_c; \\
 {[\frac{k}{k_c}]}^{1/\delta-1}, \quad for \quad k \geq k_c.
\end{array} \right.
\end{equation}
Obviously, larger $p(m)$ corresponds to higher $T$ and smaller
$E_k$, and vice versa. In this way, we compare the
rest time $\tau_k=m_k/p(m_k)$ in ZRP with the
trapping time $\tau_k=e^{\beta E_k}$ in
trap model where $\beta=1/T$~\cite{Baronchelli:2009}. It is easy to obtain: For $k< k_c$, $\beta E_k=0$;
for $k\geq k_c$, $\beta E_k=\frac{1-\delta}{\delta}log(\frac{k}{k_c})$.
Letting~$T=\frac{1}{\beta}=\delta/(1-\delta)$, then we have
\begin{eqnarray}\label{eq:trapping time}
E_k&=&0, \quad for \quad k< k_c;\nonumber \\
E_k&=&log(k)-log(k_c), \quad for \quad k\geq k_c.
\end{eqnarray}
We see that $\delta\rightarrow 0$ corresponds to
$T\rightarrow 0$, the condensation threshold $\delta_c$ corresponds to $T_c=1/(\gamma-2)$,
and $\delta\rightarrow 1$ corresponds to $T\rightarrow \infty$.
From Eq.~(\ref{eq:trapping time}) we see that larger $k$ corresponds
to larger $E_k$, i. e., deeper trap. In Ref.~\cite{Baronchelli:2009}, the average rest time
$\tau(\delta)$ is exactly the average rest time $\langle \tau\rangle=\langle k\rangle^{-1}\sum_kkP(k)\tau_k$,
indicating that the framework of ZRP is equivalent to that of
the trap model. In the thermodynamic limit of $L, N \rightarrow \infty$,
the average rest time before a hop is~$\tau(\delta)=\int_{k_0}^{k_{max}}\frac{kP(k)\tau_{k}}{\langle k
\rangle}dk$. Substituting Eq.~(\ref{eq:trapping rate}) into this equation, we have
\begin{eqnarray}\label{eq:time delay2}
\tau(\delta)&=&\frac{k_c}{\langle
k\rangle}\int_{k_0}^{k_{max}}m_{k}P(k)dk=\frac{\rho
k_c}{\langle k\rangle},
\end{eqnarray}
which depends on $\delta$ through $k_c$.
For $\delta\leq\delta_c$, substituting Eq.~(\ref{eq:kc-delta1}) into
Eq.~(\ref{eq:time delay2}) we obtain
\begin{equation}\label{eq:rest time-congestion}
\tau(\delta) \simeq \left\{\begin{array}{l}\frac{\rho}{\langle
k\rangle}{[\frac{A}{\rho(-\gamma+1+1/\delta)}]}^{\delta}
k_{max}^{1-\delta/\delta_c}, \quad for \quad \delta < \delta_c;
\\ \frac{\rho}{\langle k\rangle}{[\frac{A}{\rho}]}^{\delta}[\ln
k_{max}]^{\delta_c} , \quad for \quad \delta=\delta_c.
\end{array} \right.
\end{equation}
For $\delta_c<\delta<1$, the crossover degree $k_c$ can be obtained
from Eq.~(\ref{eq:kc-delta2}). Especially, for the case of
$\delta=1$, we have $k_c=\langle k\rangle/\rho$ when
$m_{k_c}=1$~\cite{Tang:2006}. Substituting it into
Eq.~(\ref{eq:time delay2}) we have $\tau=1$, which is consistent
with the case of random walk.

\begin{figure}
\epsfig{figure=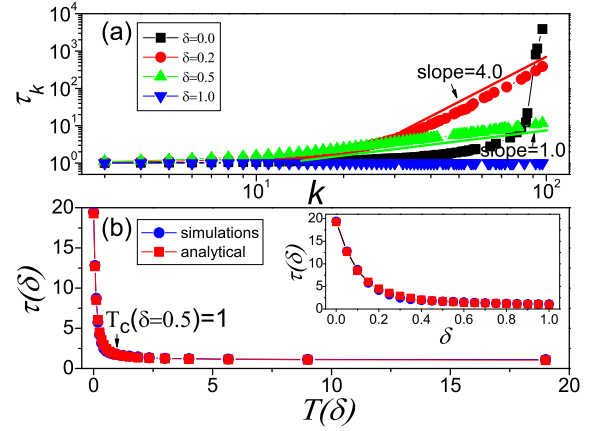,width=1\linewidth}\caption{(color online).
(a) The rest time~$\tau_k$~vs degree~$k$~for~$\delta=0.0,0.2,0.5,1.0$. The solid
lines represent the slopes $s=1/\delta-1=1.0$ and $4.0$, respectively.
(b) The average rest time $\tau(\delta)$ versus the temperature
$T(\delta)$. The ``squares" represent the numerical simulations and
the ``circles" represent the theoretical results according to Eq. (\ref{eq:time delay2}).
The inset shows how $\tau(\delta)$ changes with~$\delta$. }
\label{Fig:rest time}
\end{figure}

To confirm the theoretical results, we make numerical simulations
in an uncorrelated configuration model
(UCM)~\cite{Catanzaro:2005}. We first construct an UCM network with size $L=10^4$ and degree
distribution $P(k)\sim Ak^{-3}$, where $A\approx 13$. Other parameters of this network are
$\langle k\rangle\approx 5, k_0=3$, and $k_{max}=97$, where
$k_0$ and $k_{max}$ denote the minimum and maximum degree of the
network, respectively. We set the particle density $\rho=1$ and
let the particles hop in the network.
In the stationary state, we calculate the mean rest time~$\tau_k$~at the nodes with the
same degree $k$, and figure out their average~$\tau$. The results shown in Fig.~\ref{Fig:rest time} are
consistent with the theoretical predictions in Eqs.~(\ref{eq:trapping rate})~and~(\ref{eq:time delay2}).
As shown in Fig.~\ref{Fig:rest time}~(b), the average rest time
$\tau(\delta)$ decreases significantly with the increase of
$\delta$ for $\delta<\delta_c$, indicating that the diffusion is remarkably slowed down
in the condensation phase.

In the stationary state, a normalized particle distribution~$P_{eq}(k)$, which is defined as the probability
for a particle to be in any node with degree~$k$, is given by~$P_{eq}(k)=P(k)m_k/\rho$,
and thus
\begin{equation}\label{eq:particle distribution}
P_{eq}(k) \sim \left\{\begin{array}{l}\frac{Ak^{-\gamma+1}}{\rho k_c}, \quad for \quad k < k_c;
\\ \frac{Ak^{-\gamma+1/\delta}}{\rho k_{c}^{1/\delta}}, \quad for \quad k\geq k_c.
\end{array} \right.
\end{equation}

\begin{figure}
\epsfig{figure=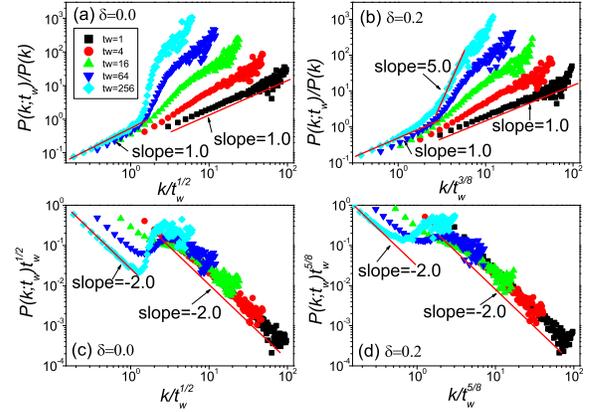,width=1\linewidth}\caption{(color online).
The scaling form of non-equilibrium distribution with~$\delta=0.0,0.2$
in the transient state. Data in both (a) and (b) obey the scaling in Eq. (\ref{eq:pktw_scaling}),
those in both (c) and (d) obey the scaling in Eq. (\ref{eq:pktw_scaling2}).}
\label{Fig:scaling}
\end{figure}

On the other hand, the relaxation dynamics in the condensation phase
is an interesting problem~\cite{Noh:2005b,Tang:2006}.
Here we investigate how~$P(k,t_w)$ converges to~$P_{eq}(k)$ in the transient state.
Until now, there is no exact theory for the relaxation dynamics of the
ZRP, and almost all the studies have been investigated by Monte Carlo simulations.
In the transient period, the simulated results
stir up one conjecture that the relaxation dynamics has a hierarchical nature~\cite{Noh:2005b,Baronchelli:2009}:
At first, subnetwork with small degree nodes is stable,
and then larger degree regions progressively equilibrate.
In fact, the nodes with small degrees correspond to shallow minima,
which take less time to explore, while the nodes with large degrees are deep
traps which take longer time to equilibrate. At time $t_w$,
one can suppose that the nodes with $k\leq k_w$ are
``at equilibrium'', while the nodes with $k\gg k_w$ are still in the random walk regime.
It turns out that the particle distribution behaves in each
regime as
\begin{equation}\label{eq:relaxation distribution}
P(k;t_w) \sim \left\{
 \begin{array}{ll}
  k^{-\gamma+1} & \mbox{\quad for \quad} k \lesssim k_v; \\
  k^{-\gamma+1/\delta} & \mbox{\quad for \quad} k_v \lesssim k \lesssim k_w; \\
  k^{-\gamma+1} & \mbox{\quad for \quad} k_w \lesssim k. \
 \end{array}
\right.
\end{equation}
Like $k_c$ in the stationary state, $k_v$~plays
the role of the crossover degree scale in a subnetwork with the largest degree~$k_w$ and the network size~$L'\sim k_w^{\delta_c}$,
that is,
\begin{equation}\label{kv_kw}
k_v \sim \left\{
\begin{array}{ll}
k_w^{1-\delta/\delta_c} & \mbox{\quad for\quad} \delta<\delta_c; \\ [2mm]
(\ln k_w)^{\delta_c} &      \mbox{\quad for\quad} \delta=\delta_c.
\end{array}\right.
\end{equation}
Considering that the total time $t_w$ is the sum of the trapping times of the visited
nodes, which is dominated by the longest one $\tau_k$ from Eq.~(\ref{eq:trapping rate}),
we obtain $k_v \sim t_w^{(\delta_c-\delta)/(1-\delta)}$ and $k_w \sim t_w^{\delta_c/(1-\delta)}$
for~$\delta<\delta_c$~in the looped networks. Figure~\ref{Fig:scaling}~(a)~and~(b)~show indeed that the
whole non-equilibrium distribution can be cast into the scaling form
\begin{equation}\label{eq:pktw_scaling}
P(k;t_w)= P(k) F\left(k / t_w^{(\delta_c-\delta)/(1-\delta)}\right),
\end{equation}
where $F$ is a scaling function such that $F(x)$ displays the scaling like~Eq.~(\ref{eq:occupation})
at small $x$, and $F(x)\sim x$ at large $x$. In addition, as shown in Fig.~\ref{Fig:scaling}~(c)~and~(d),
the whole non-equilibrium distribution also displays the scaling
\begin{equation}\label{eq:pktw_scaling2}
P(k;t_w)= t_w^{-\delta_c/(1-\delta)} G\left(k / t_w^{\delta_c/(1-\delta)}\right),
\end{equation}
where $G(x)\sim x^{1-\gamma}$ at large $x$.

This evolution takes place until the nodes with the largest degree $k_{max}$,
equilibrate. For an uncorrelated scale-free network, $k_{max} \sim
L^{1/2}$ so that the equilibration time is
\begin{equation}\label{eq:Tau}
T_{eq} \sim k_{max}^{(1-\delta)/\delta_c} \sim
L^{1-\delta}.
\end{equation}
It is consistent with the results in Ref.~\cite{Noh:2005b,Tang:2006}.

\begin{figure}
\epsfig{figure=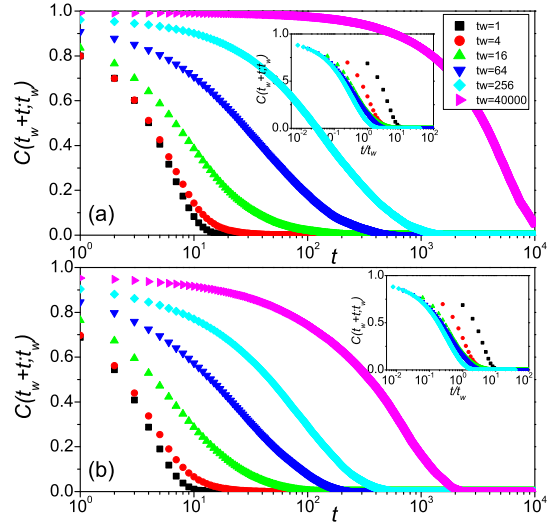,width=1.1\linewidth}\caption{(color online).
$C(t_w+t, t_w)$~vs~$t$ with~$\delta=0.0$ (a) and $\delta=0.2$ (b) in the condensation phase.
The insets show $C(t_w+t, t_w)$~vs~$t/t_w$.}
\label{Fig:two-time correlation}
\end{figure}

The evolution of $P(k;t_w)$ in the condensation phase corresponds to the
aging dynamics of the system, which has a hierarchical nature.
This dynamics is also investigated through a
two-time correlation function $C(t_w+t, t_w)$ between the states of
the system at times $t_w$ and $t_w+t$, defined as the
the average probability that a particle has not changed trap between
$t_w$ and $t_w+t$~\cite{Bouchaud:1992}: this amounts to considering that the correlation is
$1$ within one trap and $0$ between distinct traps.
The probability that a walker remains in trap $i$
longer than $t$ is simply given by $\exp(-t/\tau_i)$, so that
\begin{equation}\label{eq:two-time correlation}
C(t_w+t, t_w) = \int dk \; P(k;t_w) e^{-t/\tau_k}.
\end{equation}
In Fig.~\ref{Fig:two-time correlation}, simulations show that
the closer ZRP systems approach the stationary state, the more slowly
particles diffuse, which seems like the critical slowing down in phase transition.
As the stationary state is approached, more and more particles are trapped in hubs,
and it's very difficult for these particles to escape from hubs.
Thus, the correlation length (i. e., the characteristic time for the two-time correlation to disappear)
becomes longer and longer~\cite{Tang:2006}. For ZRP dynamics in uncorrelated scale-free networks,
it is obvious that the correlation function doesn't obey the simple aging
$C(t_w+t, t_w)=g(t/t_w)$ because of the increasingly particle attraction
in the transient state.

Aging properties of the system can also be measured through the average
escape time $t_{esc}(t_w)$ required by the random walker to escape from the
node it occupies at time $t_w$. We define $t_{esc}=\langle t'\rangle-t_w$, where
$t'>t_w$ is the time of the first jump performed by the walker after
$t_w$, which gives $t_{esc}(t_w) = \int dk \; \tau_k P(k;t_w)$.  For
small $t_w$ with respect to the equilibration time, $t_{esc}$ is
growing due to the evolution of $P(k;t_w)$. After a long time,
$m_k(t_w) \to m_k^{eq}$ in any finite system, and then we have
$t_{esc}^{eq}=  \int_{k_0}^{k_{max}} dk \; \frac{P(k)m_k\tau_k}{\rho}$.
Substituting Eq.~(\ref{eq:occupation})~and~(\ref{eq:trapping rate})
into this equation, we can numerically calculate $t_{esc}^{eq}$
for the different $\delta$, $t_{esc}^{eq}=\int_{k_0}^{k_{c}}\frac{P(k)m_{k}}{\rho}dk+\int_{k_c}^{k_{max}}\frac{P(k)m_{k}^{2-\delta}}{\rho}dk$.
For $\delta \leq \delta_c$, the system is in the condensation phase.
The nodes with $k_c \leq k \leq k_{max}$ have the capacity to
accommodate most particles. Therefore, we have~$t_{esc}^{eq}\simeq\int_{k_c}^{k_{max}}\frac{P(k)m_{k}^{2-\delta}}{\rho}dk$,
and obtain the scaling as follows
\begin{equation}\label{eq:escape time4}
t_{esc}^{eq}\sim{k_{max}}^{(\gamma-1)(1-\delta)}\sim L^{1-\delta}.
\end{equation}

\begin{figure}
\epsfig{figure=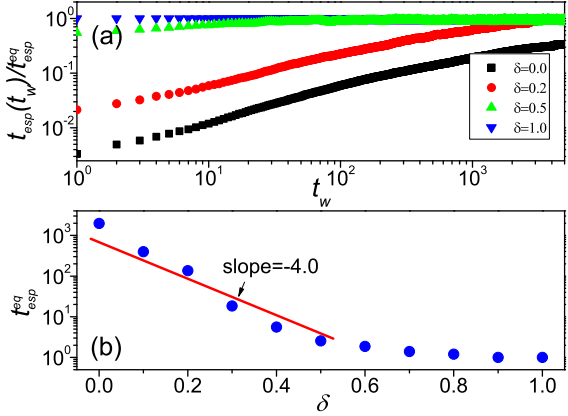,width=1\linewidth}\caption{(color online).
The average escape time~$t_{esc}$~in both the transient state and the stationary state.
(a)~$t_{esc}(t_w)/t_{esc}^{eq}$~vs~$t_w$ for $\delta=0.0,0.2,0.5,1.0$ in the transient state.
(b)~$t_{esc}^{eq}$~vs~$\delta$ where the solid line represents the slope
$s=-log_{10}L=-4.0$ from Eq.~(\ref{eq:escape time4}).}
\label{Fig:escape time}
\end{figure}

In summary, we have set up the relationship between ZRP and traps model, i. e.,~$T=\delta/(1-\delta)$,
and then investigated the relaxation dynamics of ZRP in the framework of traps model.
The particle interaction, especially the resulted condensation, is found to
significantly influence the dynamics. In the stationary state, a rest time
$\tau(\delta)=\rho k_c/\langle k\rangle$ is caused by the attraction among
particles. In the transient state, a hierarchical nature of the
aging dynamics for~$\delta<\delta_c$~is revealed by means of the
scaling analysis of particle distribution. In addition, the equilibration
time has the scaling~$T_{eq}\sim L^{1-\delta}$. Moreover, the slow dynamics
in both the transient state and the stationary state have been compared
by two important statistical measures, the two-time correlation function
and the average escape time. Both simulations and scaling analysis show that
the closer ZRP systems approach the stationary state, the more slowly
particles diffuse. At a long time, the average escape time in the condensation
phase displays the scaling~$t_{esc}^{eq}\sim L^{1-\delta}$. It is expected
that the present work will be useful for understanding the slow dynamics
of condensation in the real world.

\acknowledgments
This work is supported by the NNSF of China (Grants No. 11105025, 61103109), China Postdoctoral
Science Foundation (Grant No. 20110491705), the Specialized Research
Fund for the Doctoral Program of Higher Education (Grant No. 20110185120021), and
the Fundamental Research Funds for the Central Universities (Grant No. ZYGX2011J056).

\end{document}